\documentclass[onefignum, onetabnum,onealgnum, onethmnum]{siamonline190516}
\usepackage{lipsum}

\usepackage{endnotes}

\usepackage{titlesec}
\usepackage{titletoc}

\titleclass{\task}{straight}[\section]
\newcounter{task}
\renewcommand{\thetask}{\arabic{task}}
\titleformat{\task}[hang]
    {\normalfont\LARGE\bfseries}{Task \thetask:}{1em}{}

\titleformat*{\task}{\color{header1}\bfseries}

\titlecontents{task}
              [3.8em] 
              {}
              {\contentslabel{2.3em}}
              {\hspace*{-2.3em}}
              {\titlerule*[1pc]{.}\contentspage}

\titlespacing*{\section}{0ex}{1ex}{1ex}
\titlespacing*{\subsection}{0ex}{1ex}{1ex}
\titlespacing*{\paragraph}{0ex}{1ex}{1ex}
\titlespacing*{\subparagraph}{0pt}{1ex}{1ex}
\titlespacing*{\task}{0em}{1ex}{1ex}

\usepackage[sort&compress,comma,square,numbers]{natbib}

\usepackage{amsfonts,amsmath,amssymb}
\usepackage{bbm}
\usepackage{xspace}

\usepackage{multicol}
\usepackage{enumitem}

\usepackage{hhline}

\usepackage{comment}
\usepackage{todonotes}
\RequirePackage{ulem}

\usepackage{booktabs}
\usepackage{titlesec}
\usepackage{fancyhdr}
\usepackage{fullpage}

\RequirePackage{titlesec}
\RequirePackage[font={small},{singlelinecheck=false}]{caption}
\usepackage[T1]{fontenc}
\usepackage[scaled]{helvet}
\usepackage[scaled=1.10, med]{zlmtt}

\usepackage{algorithm}
\usepackage{algpseudocode}

\usepackage[super]{nth}
\usepackage{mdframed}
\mdfdefinestyle{MyFrame}{%
    outerlinewidth=6pt,
    roundcorner=20pt,
    innertopmargin=\baselineskip,
    innerbottommargin=\baselineskip,
    innerrightmargin=20pt,
    innerleftmargin=20pt 
    }

\providecommand{\sct}[1]{{\sc \texttt{#1}}}

\newcommand{\Mgc}{\sct{Mgc}}

\newcommand{\Hhg}{\sct{Hhg}}
\newcommand{\Dcorr}{\sct{Dcorr}}

\newcommand{\Hsic}{\sct{Hsic}}

\newcommand{\RV}{\sct{RV}}
\newcommand{\CCA}{\sct{Cca}}

\providecommand{\ve}[1]{\boldsymbol{#1}}
\providecommand{\ma}[1]{\boldsymbol{#1}}

\providecommand{\mc}[1]{\mathcal{#1}}
\providecommand{\mb}[1]{\boldsymbol{#1}}

\providecommand{\mt}[1]{\widetilde{#1}}

\providecommand{\mv}[1]{\vec{#1}}
\providecommand{\mh}[1]{\hat{#1}}

\providecommand{\mtb}[1]{\widetilde{\boldsymbol{#1}}}

\newcommand{\conv}{\rightarrow}

\newcommand{\mcE}{\mathcal{E}}

\newcommand{\defeq}{\coloneqq}

\newcommand{\PP}{\mathbb{P}}

\DeclareMathOperator{\R}{R} 
\DeclareMathOperator{\Y}{\mathbf{Y}}

\DeclareMathOperator{\X}{\mathbf{X}}

\newcommand{\bth}{\ve{\theta}}

\newcommand{\thetn}{\ve{\theta}}
\newcommand{\thet}{\thetn}

\newcommand{\FastDcorr}{\sct{Fast Dcorr}}
\newcommand{\Mmd}{\sct{Mmd}}
\newcommand{\Disco}{\sct{Disco}}
\newcommand{\Energy}{\sct{Energy}}
\newcommand{\Kmerf}{\sct{Kmerf}}

\newcommand{\Manova}{\sct{Manova}}
\newcommand{\Hotelling}{\sct{Hotelling}}
\newcommand{\MaxMargin}{\sct{MaxMargin}}

\newcommand{\mgc}{\texttt{hyppo}}
\newcommand{\Python}{\texttt{Python}}
\newcommand{\Rlang}{\texttt{R}}

\newcommand{\ksample}{\textit{k}-sample}

\usepackage{calc}
\usepackage{pythonhighlight}

\title{\mgc: A Multivariate Hypothesis Testing Python Package}

\author{%
    Sambit Panda$^1$,
    Satish Palaniappan$^1$,
    Junhao Xiong$^1$,
    Eric W. Bridgeford$^2$,
    Ronak Mehta$^1$,
    Cencheng Shen$^3$, and
    Joshua T. Vogelstein$^{1,4}$
    \thanks{Corresponding author: 
    \href{mailto:jovo@jhu.edu}{jovo@jhu.edu}
        $^1$ Johns Hopkins University, Baltimore, MD
        $^2$ Johns Hopkins School of Public Health, Baltimore, MD
        $^3$ University of Delaware, Newark, DE
        $^4$ Progressive Learning
    } 
}

\begin{document}

\maketitle

\begin{abstract}
We introduce \mgc, a unified library for performing multivariate hypothesis testing, including independence, two-sample, and \ksample~testing.
While many multivariate independence tests have \Rlang~packages available, the interfaces are inconsistent and most are not available in \Python. \mgc~includes many state of the art multivariate  testing procedures. The package is easy-to-use and is flexible enough to enable future extensions. 
The documentation and all releases are available at \url{https://hyppo.neurodata.io}.
\end{abstract}


\section{Introduction}
\label{sec:intro}

Examining and identifying relationships between sets of high-dimensional variables is critical to advance understanding and planning of future numerical and physical experiments. Hypothesis testing enables formally testing models to identify such discrepancies.  

Many correlation measures have been proposed to solve the problem of independence testing, such as Pearson's correlation \citep{pearson1895vii}, but many are unsuited to detect nonlinear and high-dimensional dependence structures within data. Recently, several statistics have been proposed that operate well on high-dimensional (potentially non-Euclidean) data, such as distance correlation \citep{szekely2009brownian, szekely2013distance, szekely2007measuring, lyons2013distance} and Hilbert-Schmidt independence criterion \citep{gretton2010consistent, gretton2005kernel, muandet2017kernel}, which are actually exactly equivalent in \citet{sejdinovic2013equivalence, shen2018exact}.
Heller, Heller and Gofrine proposed another nonparametric independence test with particularly high power in certain nonlinear relationships~\cite{heller2012consistent}.
Multiscale Graph Correlation is a test that has demonstrated higher statistical power on many multivariate, nonlinear, and structured data when compared to other independence tests \citep{shen2019distance, lee2019network, vogelstein2019discovering}, which combines and extends the nearest neighbors and energy statistics to detect underlying relationships.
For each of these tests, p-values can be calculated using a random permutation test \citep{collingridge2013primer, dwass1957modified, good2006permutation}. These tests can be modified and extended to such applications as time-series testing \citep{time1}.

To approach the problem of two-sample testing, Student's t-test \citep{student1908probable} is traditionally used, while a few nonparametric alternatives have been proposed that operate well on multivariate, nonlinear data such as Energy \citep{szekely2013energy}, and maxmimal mean discrepency \citep{gretton2012kernel}, and Heller Heller and Gorfine's test \citep{heller2012consistent}.
The two-sample testing problem can be generalized to the \ksample~testing problem and here analysis of variance (ANOVA) \citep{fisher1919xv} or its multivariate analogue multivariate ANOVA (MANOVA) \citep{bartlett1947multivariate} can be used, but these statistics either fail to, or operate poorly upon, non-Gaussian data
\citep{warne2014primer, stevens2002applied}. There are a few nonparametric alternatives to ANOVA and MANOVA, such as multivariate \ksample~Heller Heller Gorfine \citep{heller2016consistent}, and distance components (DISCO) \citep{rizzo2010disco}. Recently, \citet{shen2019exact} has shown that nonparametric distance and kernel \ksample~tests can be formulated by reducing the \ksample~testing problem to the independence testing problem.

This manuscript introduces \mgc, a hypothesis package that provides various tests with high finite-sample statistical power on multivariate and nonlinear relationships. \mgc~is a well-tested, multi-platform, \Python~3 compatible library that allows users to conduct hypothesis~tests on their data, and is also extensible enough to allow developers to easily add in their own tests. \mgc~also provides benchmarks for each of these tests by comparing statistical power over many statistical models.
The contribution of this manuscript is therefore to provide:
(1) an overview of the library and examples of how to use some of the tests in the package, and
(2) comparisons of the test statistics and wall times with similar \Rlang~packages.

\section{Library Overview}
\label{sec_lib_overview}

Inspired by the desire to allow for convenient use of these independence tests, \mgc~has been developed as a hypothesis testing package. The package structure is modeled on the \texttt{scikit-learn} and \texttt{energy} \Rlang~packages' API.
Links to source code, documentation, and tutorials can be found here: \url{https://hyppo.neurodata.io}.

\paragraph{Included Tests}
We have included a host of notable and novel hypothesis tests that we determined to be useful for the end user.
\citet{shen2018exact} has shown that distance and kernel methods are equivalent and thus, we have one implementation that is able to perform both with a proper bijective transformation. We have implemented \ksample~tests as specified in \citet{shen2019exact} and every algorithm in the following list can be used as a two-sample or \ksample~test this way.
The included algorithms are:
\begin{itemize}[noitemsep]
    \item Multivariate generalizations of Pearson's product moment correlation: \RV~\citep{escoufier1973traitement, robert1976unifying} and Cannonical correlation analysis (\CCA) \citep{hardoon2004canonical}.
    \item Heller-Heller-Gorfine (\Hhg) \citep{heller2012consistent}: Multivariate distance-based test. 
    \item Distance correlation (\Dcorr): Both biased \citep{szekely2007measuring} and unbiased \citep{szekely2014partial} and a fast $\mathcal{O}(n \log n)$ variant that runs on Euclidean one-dimensional data \citep{CHAUDHURI201915}.
    \item Hilbert-Schmidt independence criterion (\Hsic): Both biased and unbiased \citep{gretton2008kernel} kernel-based statistics. A chi-square fast statistic is also implemented \citep{shen2020chisquare}.
    \item Multiscale graph correlation (\Mgc) \citep{vogelstein2019discovering}: An independence tests that combines \textit{k}-nearest neighbors and energy statistics. Recently, \Mgc~has been accepted into \href{https://docs.scipy.org/doc/scipy/reference/generated/scipy.stats.multiscale_graphcorr.html}{\texttt{scipy.stats}} and this implementation wraps the \texttt{scipy} implementation.
    \item Friedman Rafsky \citep{friedman1979multivariate}: A tree-based two sample test.
    \item dHsic \citep{pfister2018kernel}: A d-variate independence test based on~\Hsic
    \item Multivariate analysis of variance (\Manova) \citep{carey1998multivariate, warne2014primer} and Hotelling $T^2$ (\Hotelling) \citep{hotelling1992generalization}.
    \item Maximum mean discrepancy (\Mmd) \citep{gretton2012kernel}: A kernel two-sample test.
    \item \Energy~\citep{szekely2013energy}: A distance two-sample test.
    \item Distance components (\Disco) \citep{rizzo2010disco}: A distance-based \ksample~test.
    \item Smooth CF Test \citep{chwialkowski2015fast}: A test using analytic analogues of characteristic functions.
    \item Mean Embedding Test \citep{chwialkowski2015fast}: A test based on the analytical mean embeddings between two distributions.
    \item KSample~\Hhg~\citep{heller2016multivariate}: A $k$-sample test for~\Hhg.
    \item Fast Conditional Independence Test \citep{chalupka2018fast}: A fast, nonparametric conditional independence test.
    \item Kernel Conditional Independence Test \citep{zhang2012kernel}: An efficient, kernel conditional independence test.
    \item Finite Set Stein Discrepancy \citep{jitkrittum2017linear}: A linear time kernel goodness of fit test.
    \item Discriminability test \citep{wang2020statistical}: A highly accurate and powerful discriminability test.
    \item Partial \Dcorr~\citep{szekelyPartialDistanceCorrelation2014a}: A method to perform conditional independence testing using distance correlation.
    \item Conditional \Dcorr~\citep{wang2015conditional}: Conditional independence testing using Dcorr with strong theoretical properties.
    \item LjungBox \citep{ljung1978measure}: Tests if groups of autocorrelations of time series are different from 0.
\end{itemize}
A number of algorithms have been implemented that lack an open source implementation elsewhere. These include:
\begin{itemize}[noitemsep]
    \item Kernel mean embedding random forest (\Kmerf) \citep{shen2020learning}: A kernel test that leverages random forest kernel induced similarity matrix to generate a test statistic.
    \item Fast Implementations of \Dcorr~(\FastDcorr) \citep{shen2020hd}: An approximation to \Dcorr~when calculating the p-value.
    \item Universally consistent $k$-sample tests via independence testing \citep{shen2019exact}: Transforms the \ksample~testing problem into the independence testing problem and then uses non-parametric independence tests from \mgc.
    \item Time-series \Mgc~and \Dcorr: Applying \Mgc~and \Dcorr~to time-series data.
    \item Maximal Margin Correlation (\MaxMargin) \citep{shen2020hd}: A highly accurate formulation of independence tests in high dimensions with minimal added computational complexity.
\end{itemize}

%

\paragraph{Structure of \mgc}
The modules of \mgc~are as follows: \texttt{independence}, \texttt{d\_variate}, \texttt{ksample},
\texttt{discrim}, \texttt{kgof}
\texttt{time\_series}, and \texttt{tools}.
Each test within \mgc~contains a \pyth{.test} method which the user runs that returns at least a statistic and p-value in all cases, and parallelization of the permutation tests are performed using \texttt{joblib}.
Test statistic code is compiled and cached with \texttt{numba}, which is a just-in-time (JIT) compiler.
\texttt{tools} contains common helper functions, simulations, and finite-sample statistical power functions to test differences between each test in \mgc.


\section{Benchmarks}
\label{sec_benchmarks}

\begin{table}
    \begin{tabular}{cc}
        \raisebox{2ex - \height}
        {\includegraphics[width=8cm]{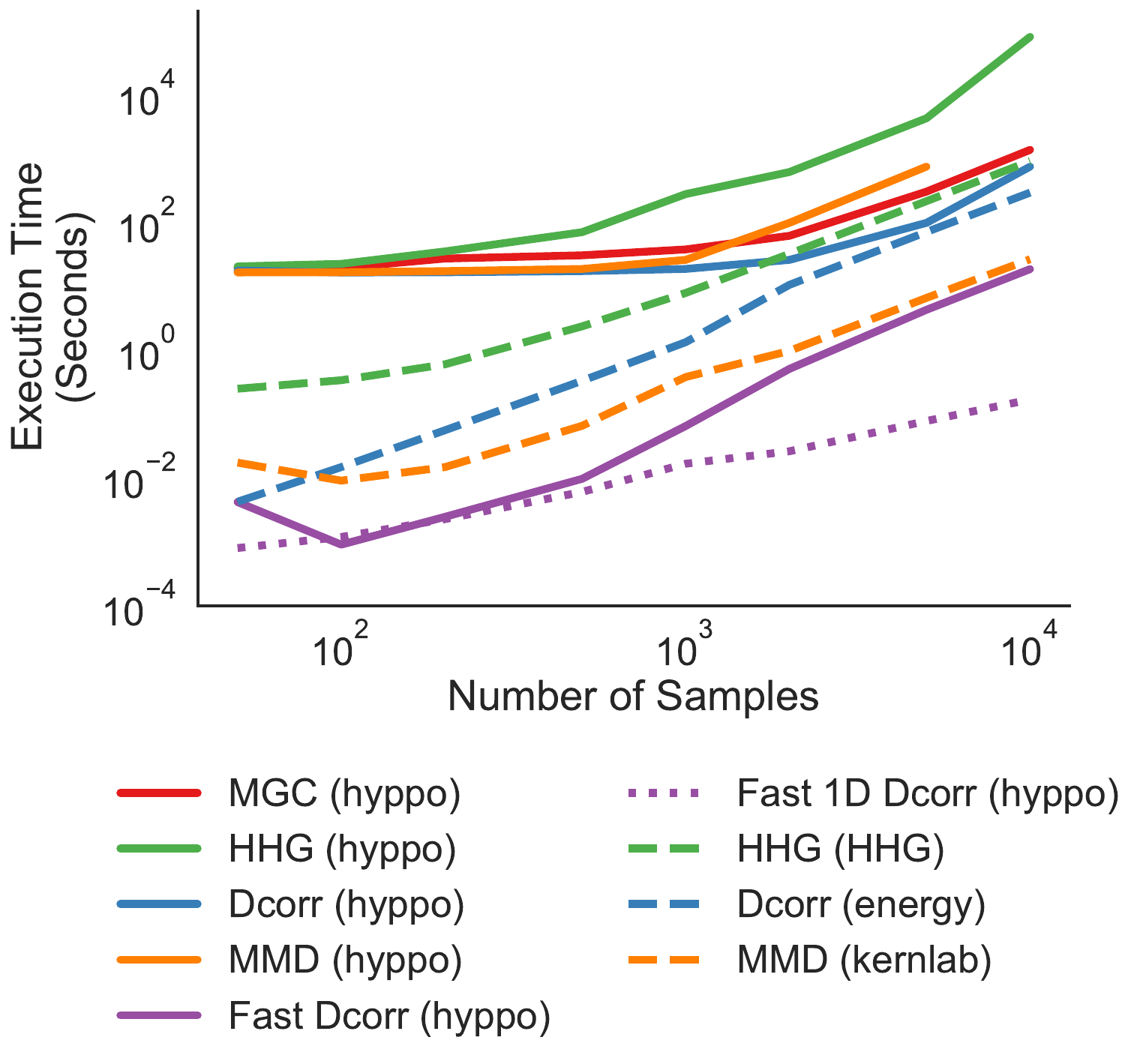}} &
        \raisebox{2ex - \height}
        {\includegraphics[width=6cm]{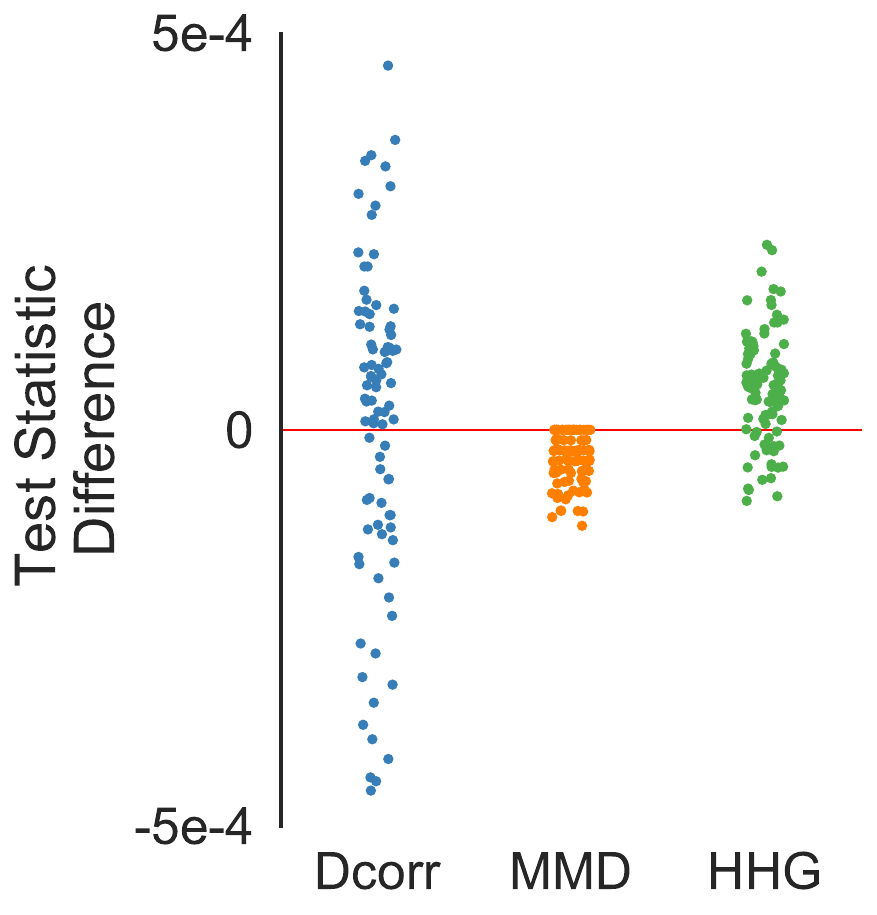}}
    \end{tabular}
    \captionof{figure}{Benchmarks of \mgc~implementations against corresponding \Rlang~implementations for tests in the independence testing module. Average wall times (over 3 repetitions) (left) are shown for \Dcorr~in \texttt{energy}, \Mmd~in \texttt{kernlab}, and \Hhg~in \texttt{Hhg} as compared against \mgc~implementations of \Mgc, \Hhg, \Dcorr, \Mmd, and \FastDcorr. Test statistic comparisons (right) between \Dcorr, \Mmd, and \Hhg~in \mgc~are compared against their respective reference \Rlang~implementations. Test statistics are nearly identical for each implementation.}
    \label{fig_1}
\end{table}

\paragraph{Wall Time Comparisons}

Figure 
\ref{fig_1}a 
shows the computational efficiency of \mgc's implementations against existing implementations in commonly used \Rlang~packages---specifically \texttt{energy} \citep{rizzo2018energy}, \texttt{kernlab} \citep{karatzoglou2004kernlab}, and \texttt{HHG} \citep{brill2019hhg}. 
When comparing performance, wall times are averages of p-value computations (1000 replications when permutation tests are used) 3 trials  calculated on a univariate noisy linear simulation  with number of samples increasing from 50 to 10,000. All computations were performed on an Ubuntu 18.04.3 LTS system with access to 96 cores. 
When sample sizes are above a few hundred, all algorithms achieve approximately quadratic times, with different slopes.
\Hhg~was the slowest as expected, though had comparable speeds to the other algorithms at low sample sizes.
\Mgc~and \Dcorr~are next, and still only requires tens of minutes to run when sample sizes are around 10,000.
At low sample sizes, the \texttt{energy}~package's \Dcorr~is faster than \texttt{kernlab}'s implementation of \Mmd~(\Dcorr~is equivalent to \Mmd~for all finite sample sizes~\cite{shen2018exact})
even at a sample size of 10,000.
\mgc's \FastDcorr, which uses a fast statistic \citep{CHAUDHURI201915} and p-value approximation \citep{shen2020chisquare} is the fastest, even though both \texttt{energy} and \texttt{kernlab} both use highly optimized C++ versions.

\paragraph{Implementation Validation}

Next, we verify that  \mgc's  test statistics are equivalent to existing \Rlang~implementations of the tests. 
Specifically, \mgc's implementations were compared to: \Dcorr~from the \texttt{energy} package \citep{rizzo2018energy}.
\Mmd~from the \texttt{kernlab} package \citep{karatzoglou2004kernlab},
and \Hhg~from the \texttt{HHG} package \citep{brill2019hhg}.
The evaluation uses a spiral simulation with 1000 samples and 2 dimensions for each test and compares test statistics over 20 repetitions. Figure~\ref{fig_1}b shows the difference between the \mgc~implementation of the independence test and the respective \Rlang~package implementation of the independence test. 
Test statistics are nearly equivalent for each implementation.

\section{Conclusion}
\label{sec_conclusion}
\mgc~is an extensive and extensible open-source \Python~package for multivariate hypothesis testing. 
As \mgc~continues to grow and add functionality, it will enhance tools scientists use when determining relationships within their investigations.

\paragraph{Acknowledgements}
This work is graciously supported by the Defense Advanced Research Projects Agency (DARPA) Lifelong Learning Machines program through contract FA8650-18-2-7834, the National Institute of Health awards RO1MH120482 amd T32GM119998, and the National Science Foundation award DMS-1921310.  The authors would like to acknowledge the NeuroData Design class and the NeuroData lab at Johns Hopkins University for helpful feedback.

\clearpage

\bibliographystyle{unsrtnat}
\bibliography{references}

\begin{thebibliography}{53}
\providecommand{\natexlab}[1]{#1}
\providecommand{\url}[1]{\texttt{#1}}
\expandafter\ifx\csname urlstyle\endcsname\relax
  \providecommand{\doi}[1]{doi: #1}\else
  \providecommand{\doi}{doi: \begingroup \urlstyle{rm}\Url}\fi

\bibitem[Pearson(1895)]{pearson1895vii}
Karl Pearson.
\newblock Note on regression and inheritance in the case of two parents.
\newblock \emph{Proceedings of the Royal Society of London}, 58\penalty0
  (347-352):\penalty0 240--242, 1895.

\bibitem[Sz{\'e}kely and Rizzo(2009)]{szekely2009brownian}
G{\'a}bor~J Sz{\'e}kely and Maria~L Rizzo.
\newblock Brownian distance covariance.
\newblock \emph{The Annals of Applied Statistics}, 3\penalty0 (4):\penalty0
  1236--1265, 2009.

\bibitem[Sz{\'e}kely and Rizzo(2013{\natexlab{a}})]{szekely2013distance}
G{\'a}bor~J Sz{\'e}kely and Maria~L Rizzo.
\newblock The distance correlation t-test of independence in high dimension.
\newblock \emph{Journal of Multivariate Analysis}, 117:\penalty0 193--213,
  2013{\natexlab{a}}.

\bibitem[Sz{\'e}kely et~al.(2007)Sz{\'e}kely, Rizzo, and
  Bakirov]{szekely2007measuring}
G{\'a}bor~J Sz{\'e}kely, Maria~L Rizzo, and Nail~K Bakirov.
\newblock Measuring and testing dependence by correlation of distances.
\newblock \emph{The Annals of Statistics}, 35\penalty0 (6):\penalty0
  2769--2794, 2007.

\bibitem[Lyons(2013)]{lyons2013distance}
Russell Lyons.
\newblock Distance covariance in metric spaces.
\newblock \emph{The Annals of Probability}, 41\penalty0 (5):\penalty0
  3284--3305, 2013.

\bibitem[Gretton and L\'{a}szl\'{o}(2010)]{gretton2010consistent}
Arthur Gretton and Gy\"{o}rfi L\'{a}szl\'{o}.
\newblock Consistent nonparametric tests of independence.
\newblock \emph{Journal of Machine Learning Research}, 11\penalty0
  (Apr):\penalty0 1391--1423, 2010.

\bibitem[Gretton et~al.(2005)Gretton, Herbrich, Smola, Bousquet, and
  Sch{\"o}lkopf]{gretton2005kernel}
Arthur Gretton, Ralf Herbrich, Alexander Smola, Olivier Bousquet, and Bernhard
  Sch{\"o}lkopf.
\newblock Kernel methods for measuring independence.
\newblock \emph{Journal of Machine Learning Research}, 6\penalty0
  (Dec):\penalty0 2075--2129, 2005.

\bibitem[Muandet et~al.(2017)Muandet, Fukumizu, Sriperumbudur, and
  Sch{\"o}lkopf]{muandet2017kernel}
Krikamol Muandet, Kenji Fukumizu, Bharath Sriperumbudur, and Bernhard
  Sch{\"o}lkopf.
\newblock Kernel mean embedding of distributions: A review and beyond.
\newblock \emph{Foundations and Trends{\textregistered} in Machine Learning},
  10\penalty0 (1-2):\penalty0 1--141, 2017.

\bibitem[Sejdinovic et~al.(2013)Sejdinovic, Sriperumbudur, Gretton, and
  Fukumizu]{sejdinovic2013equivalence}
Dino Sejdinovic, Bharath Sriperumbudur, Arthur Gretton, and Kenji Fukumizu.
\newblock Equivalence of distance-based and rkhs-based statistics in hypothesis
  testing.
\newblock \emph{The Annals of Statistics}, 41\penalty0 (5):\penalty0
  2263--2291, 2013.

\bibitem[Shen and Vogelstein(2021)]{shen2018exact}
Cencheng Shen and Joshua~T. Vogelstein.
\newblock The exact equivalence of distance and kernel methods in hypothesis
  testing.
\newblock \emph{AStA Advances in Statistical Analysis}, 105\penalty0
  (3):\penalty0 385--403, 2021.

\bibitem[Heller et~al.(2012)Heller, Heller, and Gorfine]{heller2012consistent}
Ruth Heller, Yair Heller, and Malka Gorfine.
\newblock A consistent multivariate test of association based on ranks of
  distances.
\newblock \emph{Biometrika}, 100\penalty0 (2):\penalty0 503--510, 2012.

\bibitem[Shen et~al.(2020)Shen, Priebe, and Vogelstein]{shen2019distance}
Cencheng Shen, Carey~E Priebe, and Joshua~T Vogelstein.
\newblock From distance correlation to multiscale graph correlation.
\newblock \emph{Journal of the American Statistical Association}, 115\penalty0
  (529):\penalty0 280--291, 2020.

\bibitem[Lee et~al.(2019)Lee, Shen, Priebe, and Vogelstein]{lee2019network}
Youjin Lee, Cencheng Shen, Carey~E Priebe, and Joshua~T Vogelstein.
\newblock Network dependence testing via diffusion maps and distance-based
  correlations.
\newblock \emph{Biometrika}, 106\penalty0 (4):\penalty0 857--873, 2019.

\bibitem[Vogelstein et~al.(2019)Vogelstein, Bridgeford, Wang, Priebe, Maggioni,
  and Shen]{vogelstein2019discovering}
Joshua~T Vogelstein, Eric~W Bridgeford, Qing Wang, Carey~E Priebe, Mauro
  Maggioni, and Cencheng Shen.
\newblock Discovering and deciphering relationships across disparate data
  modalities.
\newblock \emph{eLife}, 8:\penalty0 e41690, 2019.

\bibitem[Collingridge(2013)]{collingridge2013primer}
Dave~S Collingridge.
\newblock A primer on quantitized data analysis and permutation testing.
\newblock \emph{Journal of Mixed Methods Research}, 7\penalty0 (1):\penalty0
  81--97, 2013.

\bibitem[Dwass(1957)]{dwass1957modified}
Meyer Dwass.
\newblock Modified randomization tests for nonparametric hypotheses.
\newblock \emph{The Annals of Mathematical Statistics}, pages 181--187, 1957.

\bibitem[Good(2006)]{good2006permutation}
Phillip~I Good.
\newblock \emph{Permutation, parametric, and bootstrap tests of hypotheses}.
\newblock Springer Science \& Business Media, 2006.

\bibitem[Shen et~al.(2024{\natexlab{a}})Shen, Chung, Mehta, Xu, and
  Vogelstein]{time1}
Cencheng Shen, Jaewon Chung, Ronak Mehta, Ting Xu, and Joshua~T Vogelstein.
\newblock Independence testing for temporal data.
\newblock \emph{Transactions on Machine Learning Research}, 2024{\natexlab{a}}.
\newblock ISSN 2835-8856.
\newblock URL \url{https://openreview.net/forum?id=jv1aPQINc4}.

\bibitem[Student(1908)]{student1908probable}
Student.
\newblock The probable error of a mean.
\newblock \emph{Biometrika}, pages 1--25, 1908.

\bibitem[Sz{\'e}kely and Rizzo(2013{\natexlab{b}})]{szekely2013energy}
G{\'a}bor~J Sz{\'e}kely and Maria~L Rizzo.
\newblock Energy statistics: A class of statistics based on distances.
\newblock \emph{Journal of statistical planning and inference}, 143\penalty0
  (8):\penalty0 1249--1272, 2013{\natexlab{b}}.

\bibitem[Gretton et~al.(2012)Gretton, Borgwardt, Rasch, Sch{\"o}lkopf, and
  Smola]{gretton2012kernel}
Arthur Gretton, Karsten~M Borgwardt, Malte~J Rasch, Bernhard Sch{\"o}lkopf, and
  Alexander Smola.
\newblock A kernel two-sample test.
\newblock \emph{Journal of Machine Learning Research}, 13\penalty0
  (Mar):\penalty0 723--773, 2012.

\bibitem[Fisher(1919)]{fisher1919xv}
Ronald~A Fisher.
\newblock Xv.—the correlation between relatives on the supposition of
  mendelian inheritance.
\newblock \emph{Earth and Environmental Science Transactions of the Royal
  Society of Edinburgh}, 52\penalty0 (2):\penalty0 399--433, 1919.

\bibitem[Bartlett(1947)]{bartlett1947multivariate}
Maurice~S Bartlett.
\newblock Multivariate analysis.
\newblock \emph{Supplement to the journal of the royal statistical society},
  9\penalty0 (2):\penalty0 176--197, 1947.

\bibitem[Warne(2014)]{warne2014primer}
Russell Warne.
\newblock A primer on multivariate analysis of variance (manova) for behavioral
  scientists.
\newblock \emph{Practical Assessment, Research, and Evaluation}, 19\penalty0
  (1):\penalty0 17, 2014.

\bibitem[Stevens(2002)]{stevens2002applied}
JP~Stevens.
\newblock Applied multivariate statistics for the social sciences. lawrence
  erlbaum.
\newblock \emph{Mahwah, NJ}, pages 510--1, 2002.

\bibitem[Heller et~al.(2016)Heller, Heller, Kaufman, Brill, and
  Gorfine]{heller2016consistent}
Ruth Heller, Yair Heller, Shachar Kaufman, Barak Brill, and Malka Gorfine.
\newblock Consistent distribution-free k-sample and independence tests for
  univariate random variables.
\newblock \emph{The Journal of Machine Learning Research}, 17\penalty0
  (1):\penalty0 978--1031, 2016.

\bibitem[Rizzo et~al.(2010)Rizzo, Sz{\'e}kely, et~al.]{rizzo2010disco}
Maria~L Rizzo, G{\'a}bor~J Sz{\'e}kely, et~al.
\newblock Disco analysis: A nonparametric extension of analysis of variance.
\newblock \emph{The Annals of Applied Statistics}, 4\penalty0 (2):\penalty0
  1034--1055, 2010.

\bibitem[Panda et~al.(2024)Panda, Shen, Perry, Zorn, Lutz, Priebe, and
  Vogelstein]{shen2019exact}
Sambit Panda, Cencheng Shen, Ronan Perry, Jelle Zorn, Antoine Lutz, Carey~E.
  Priebe, and Joshua~T. Vogelstein.
\newblock High-dimensional and universally consistent k-sample tests.
\newblock \emph{arXiv preprint arXiv:1910.08883}, 2024.

\bibitem[Escoufier(1973)]{escoufier1973traitement}
Yves Escoufier.
\newblock Le traitement des variables vectorielles.
\newblock \emph{Biometrics}, pages 751--760, 1973.

\bibitem[Robert and Escoufier(1976)]{robert1976unifying}
Paul Robert and Yves Escoufier.
\newblock A unifying tool for linear multivariate statistical methods: the
  rv-coefficient.
\newblock \emph{Journal of the Royal Statistical Society: Series C (Applied
  Statistics)}, 25\penalty0 (3):\penalty0 257--265, 1976.

\bibitem[Hardoon et~al.(2004)Hardoon, Szedmak, and
  Shawe-Taylor]{hardoon2004canonical}
David~R Hardoon, Sandor Szedmak, and John Shawe-Taylor.
\newblock Canonical correlation analysis: An overview with application to
  learning methods.
\newblock \emph{Neural computation}, 16\penalty0 (12):\penalty0 2639--2664,
  2004.

\bibitem[Sz{\'e}kely et~al.(2014)Sz{\'e}kely, Rizzo,
  et~al.]{szekely2014partial}
G{\'a}bor~J Sz{\'e}kely, Maria~L Rizzo, et~al.
\newblock Partial distance correlation with methods for dissimilarities.
\newblock \emph{The Annals of Statistics}, 42\penalty0 (6):\penalty0
  2382--2412, 2014.

\bibitem[Chaudhuri and Hu(2019)]{CHAUDHURI201915}
Arin Chaudhuri and Wenhao Hu.
\newblock A fast algorithm for computing distance correlation.
\newblock \emph{Computational Statistics \& Data Analysis}, 135:\penalty0 15 --
  24, 2019.
\newblock ISSN 0167-9473.
\newblock \doi{https://doi.org/10.1016/j.csda.2019.01.016}.
\newblock URL
  \url{http://www.sciencedirect.com/science/article/pii/S0167947319300313}.

\bibitem[Gretton et~al.(2008)Gretton, Fukumizu, Teo, Song, Sch{\"o}lkopf, and
  Smola]{gretton2008kernel}
Arthur Gretton, Kenji Fukumizu, Choon~H Teo, Le~Song, Bernhard Sch{\"o}lkopf,
  and Alex~J Smola.
\newblock A kernel statistical test of independence.
\newblock In \emph{Advances in neural information processing systems}, pages
  585--592, 2008.

\bibitem[Shen et~al.(2022)Shen, Panda, and Vogelstein]{shen2020chisquare}
Cencheng Shen, Sambit Panda, and Joshua~T. Vogelstein.
\newblock The chi-square test of distance correlation.
\newblock \emph{Journal of Computational and Graphical Statistics}, 31\penalty0
  (1):\penalty0 254--262, 2022.

\bibitem[Friedman and Rafsky(1979)]{friedman1979multivariate}
Jerome~H Friedman and Lawrence~C Rafsky.
\newblock Multivariate generalizations of the wald-wolfowitz and smirnov
  two-sample tests.
\newblock \emph{The Annals of Statistics}, pages 697--717, 1979.

\bibitem[Pfister et~al.(2018)Pfister, B{\"u}hlmann, Sch{\"o}lkopf, and
  Peters]{pfister2018kernel}
Niklas Pfister, Peter B{\"u}hlmann, Bernhard Sch{\"o}lkopf, and Jonas Peters.
\newblock Kernel-based tests for joint independence.
\newblock \emph{Journal of the Royal Statistical Society Series B: Statistical
  Methodology}, 80\penalty0 (1):\penalty0 5--31, 2018.

\bibitem[Carey(1998)]{carey1998multivariate}
Gregory Carey.
\newblock Multivariate analysis of variance (manova): I. theory.
\newblock \emph{Retrieved May}, 14:\penalty0 2011, 1998.

\bibitem[Hotelling(1992)]{hotelling1992generalization}
Harold Hotelling.
\newblock The generalization of student’s ratio.
\newblock In \emph{Breakthroughs in statistics}, pages 54--65. Springer, 1992.

\bibitem[Chwialkowski et~al.(2015)Chwialkowski, Ramdas, Sejdinovic, and
  Gretton]{chwialkowski2015fast}
Kacper~P Chwialkowski, Aaditya Ramdas, Dino Sejdinovic, and Arthur Gretton.
\newblock Fast two-sample testing with analytic representations of probability
  measures.
\newblock \emph{Advances in Neural Information Processing Systems}, 28, 2015.

\bibitem[Heller and Heller(2016)]{heller2016multivariate}
Ruth Heller and Yair Heller.
\newblock Multivariate tests of association based on univariate tests.
\newblock \emph{Advances in Neural Information Processing Systems}, 29, 2016.

\bibitem[Chalupka et~al.(2018)Chalupka, Perona, and
  Eberhardt]{chalupka2018fast}
Krzysztof Chalupka, Pietro Perona, and Frederick Eberhardt.
\newblock Fast conditional independence test for vector variables with large
  sample sizes.
\newblock \emph{arXiv preprint arXiv:1804.02747}, 2018.

\bibitem[Zhang et~al.(2012)Zhang, Peters, Janzing, and
  Sch{\"o}lkopf]{zhang2012kernel}
Kun Zhang, Jonas Peters, Dominik Janzing, and Bernhard Sch{\"o}lkopf.
\newblock Kernel-based conditional independence test and application in causal
  discovery.
\newblock \emph{arXiv preprint arXiv:1202.3775}, 2012.

\bibitem[Jitkrittum et~al.(2017)Jitkrittum, Xu, Szab{\'o}, Fukumizu, and
  Gretton]{jitkrittum2017linear}
Wittawat Jitkrittum, Wenkai Xu, Zolt{\'a}n Szab{\'o}, Kenji Fukumizu, and
  Arthur Gretton.
\newblock A linear-time kernel goodness-of-fit test.
\newblock \emph{Advances in neural information processing systems}, 30, 2017.

\bibitem[Wang et~al.(2020)Wang, Bridgeford, Wang, Vogelstein, and
  Caffo]{wang2020statistical}
Zeyi Wang, Eric Bridgeford, Shangsi Wang, Joshua~T. Vogelstein, and Brian
  Caffo.
\newblock Statistical analysis of data repeatability measures, 2020.

\bibitem[Sz{\'e}kely and Rizzo(2014)]{szekelyPartialDistanceCorrelation2014a}
G{\'a}bor~J. Sz{\'e}kely and Maria~L. Rizzo.
\newblock Partial distance correlation with methods for dissimilarities.
\newblock \emph{The Annals of Statistics}, 42\penalty0 (6):\penalty0
  2382--2412, December 2014.
\newblock ISSN 0090-5364, 2168-8966.
\newblock \doi{10.1214/14-AOS1255}.

\bibitem[Wang et~al.(2015)Wang, Pan, Hu, Tian, and Zhang]{wang2015conditional}
Xueqin Wang, Wenliang Pan, Wenhao Hu, Yuan Tian, and Heping Zhang.
\newblock Conditional distance correlation.
\newblock \emph{Journal of the American Statistical Association}, 110\penalty0
  (512):\penalty0 1726--1734, 2015.

\bibitem[Ljung and Box(1978)]{ljung1978measure}
Greta~M Ljung and George~EP Box.
\newblock On a measure of lack of fit in time series models.
\newblock \emph{Biometrika}, 65\penalty0 (2):\penalty0 297--303, 1978.

\bibitem[Shen et~al.(2024{\natexlab{b}})Shen, Panda, and
  Vogelstein]{shen2020learning}
Cencheng Shen, Sambit Panda, and Joshua~T. Vogelstein.
\newblock Learning interpretable characteristic kernels via decision forests.
\newblock \emph{arXiv preprint arXiv:1812.00029}, 2024{\natexlab{b}}.

\bibitem[Shen and Dong(2024)]{shen2020hd}
Cencheng Shen and Yuexiao Dong.
\newblock High-dimensional independence testing via maximum and average
  distance correlations.
\newblock \emph{arXiv preprint arXiv:2001.01095}, 2024.

\bibitem[Rizzo and Szekely(2018)]{rizzo2018energy}
Maria Rizzo and Gabor Szekely.
\newblock \emph{energy: E-Statistics: Multivariate Inference via the Energy of
  Data}, 2018.
\newblock URL \url{https://CRAN.R-project.org/package=energy}.
\newblock R package version 1.7-5.

\bibitem[Karatzoglou et~al.(2004)Karatzoglou, Smola, Hornik, and
  Zeileis]{karatzoglou2004kernlab}
Alexandros Karatzoglou, Alex Smola, Kurt Hornik, and Achim Zeileis.
\newblock kernlab -- an {S4} package for kernel methods in {R}.
\newblock \emph{Journal of Statistical Software}, 11\penalty0 (9):\penalty0
  1--20, 2004.
\newblock URL \url{http://www.jstatsoft.org/v11/i09/}.

\bibitem[Brill and Kaufman(2019)]{brill2019hhg}
Barak Brill and Shachar Kaufman.
\newblock \emph{HHG: Heller-Heller-Gorfine Tests of Independence and Equality
  of Distributions}, 2019.
\newblock URL \url{https://CRAN.R-project.org/package=HHG}.
\newblock R package version 2.3.2.

\end{thebibliography}

\clearpage



\end{document}